\documentclass[fleqn,10pt]{wlscirep}
\UseRawInputEncoding
\usepackage[utf8]{inputenc}
\usepackage[T1]{fontenc}
\usepackage{amsmath}
\title{Feature-specific correlation of structural, optical, and chemical properties in the transmission electron microscope with hypermodal data fusion}

\author[1,*]{Thomas Thersleff}
\author[1]{Cheuk-Wai Tai}
\affil[1]{Stockholm University, Department of Materials and Environmental Chemistry, 10691 Stockholm, Sweden}

\affil[*]{thomas.thersleff@mmk.su.se}

\begin{abstract}
Modern TEM instrumentation can probe a wide range of structural, optical, and chemical properties with unprecedented resolution. However, each of these properties must be recorded in independent datasets using different detector modes with no unifying framework currently available for quantitatively mapping their relationships onto chemically distinct features, particularly in complex morphologies. Here, we tackle this challenge by proposing a data acquisition and analysis workflow called "hypermodal data fusion," describing how to directly couple an arbitrary number of highly disparate detector modes including spectroscopy and scanning diffraction and jointly analyze them for correlations. We demonstrate this concept on a random collection of anatase and rutile nanoparticles, first detailing how to use core-loss EELS to unmix the different polymorphs despite three-dimensional overlap along the beam direction and then showing how this can be used to extract polymorph-specific composition, bandgaps, and crystal structure. We conclude with a discussion on the applicability of this workflow for a broad range of materials systems.

\end{abstract}
\begin{document}

\flushbottom
\maketitle
\thispagestyle{empty}

\section*{Introduction}

Imperative to the rapid progression of nanotechnology is our ability to localize functionally important materials properties to nanoscale features in complex morphologies and heterostuctures. This involves the use of advanced spectroscopy and imaging techniques that exploit the diverse interactions of fast incident electrons with the atoms of the material under investigation. The scientific discipline of using electron optics to both customize the incident electron wavefront prior to interaction with thin samples as well as direct the resulting transmitted scattering cross-sections to specific detectors is broadly known as Transmission Electron Microscopy (TEM), and this field encompasses an enormous range of powerful nanoscale analysis methods capable of resolving atomic structures\cite{menter_direct_1956,wall_scanning_1974}.  The recent commercialization of aberration-correctors, particularly when applied to pre-specimen optics, means that it is now possible to focus the majority of incoming electrons into a probe sufficiently fine that the entire electron scattering cross-section can be localized to a lateral entrance area with a diameter of less than one {\AA}ngstr{\"o}m \cite{krivanek_towards_1999,batson_sub-angstrom_2002,pennycook_aberration-corrected_2009,brydson_aberration-corrected_2011}. This probe can be rapidly translated across the sample in a grid - a technique known as Scanning TEM (STEM)\cite{pennycook_scanning_2011} - and its position can be synchronized with multiple detectors in the microscope, often operating in parallel. Although it is currently infeasible to record the entire scattering cross section at each probe position, these detectors can be deliberately placed in positions where the portion of scattering they collect can be related to critical materials properties, permitting such experiments to directly address nanoscale scientific inquiries.

One example of how STEM techniques can be used to confront a contemporary materials science challenge comes from the field of nanoscale catalytic materials based off of titanium dioxide (TiO$_2$, also called titania). In addition to its abundance, low cost, and non-toxicity, titania exhibits a versatile crystallographic and chemical landscape that can be exploited through nanoengineering to customize the optical, electronic, and photocatalytic properties of green photocatalysts \cite{hashimoto_tio2_2005,schneider_understanding_2014}. Titania is found naturally on earth in one of three polymorphs known as rutile, anatase, and brookite that, despite identical nominal compositions, each display distinct structural, optical, and electronic properties.  Careful interface engineering of nanoscale polymorph mixtures can influence the band alignment of larger morphological features, thereby providing access to a much larger array of macroscopic photocatalytic properties than available in single-phase systems \cite{hurum_explaining_2003,scanlon_band_2013,zhang_importance_2008,li_solid-solid_2007,zheng_hollow_2018,snair_converting_2020}. Understanding these materials requires not only techniques that can quantify relationships between their structural, optical, and chemical properties, but that can moreover map these properties to nanosized morphological features. At present, this second requirement can is most readily met with electron microscopy, making it an essential tool for driving this field forward.

As illustrated in figure \ref{fig:DFschematic} below, the critical materials properties for most materials systems (including titania) are usually individually extracted from separate detectors in the electron microscope that each collect different scattering cross-sections. Some of these properties, such as composition and electronic structure, can be collected in parallel, while others, such as optical properties and crystallographic orientation, require configuring a single detector in different collection modes. In both cases, the experimentalist is left with multiple independent datasets. Moreover, the properties of interest are often mixed since the electrons transmit through a volume of material that may consist of multiple phases. This poses an enormous challenge to the subsequent analysis, as addressing the scientific question requires each dataset to be analyzed individually and then somehow related to the other datasets. Traditionally, this has been achieved by mapping the materials properties in real space and then qualitatively comparing the results \cite{thersleff_single-pass_2019}. However, not only does this approach fail to designate individual properties to independent morphological features, it is incapable of producing a quantitative relationship describing how one materials property directly relates to another one, which is the ultimate underlying goal of the experiment.  Thus, using titania as an example, while the scientific question may endeavor to isolate the optical properties of an individual polymorph in an agglomerate of overlapping particles, at present, there is no well-established experimental framework that can achieve this.

In this report, we demonstrate how an experimental and analytical framework we are developing called "hypermodal data fusion" can directly address these challenges.  The word "hypermodal" is intended to reflect the disparate nature of the datasets needed to extract the desired optical, structural, and chemical properties, which include both hyperspectral datasets as well as detector modes capable of acquiring scanning diffraction patterns. "Data fusion" refers to the established technique of coupling multiple datasets via shared factors to uncover correlations between them\cite{cocchi_data_2019}.  We have previously shown how data fusion can be applied to simultaneously-acquired hyperspectral data to "extend" the spectral range of experiments, leading to greatly improved chemical mapping capabilities \cite{thersleff_soot_2021, merkl_plasmonic_2021}.  It can also be complemented with blind-source separation techniques to permit dissection of complex functional materials into their constituent morphological, compositional, and electronic components \cite{thersleff_dissecting_2020}.  Here, we extend this to include local structure measurements by additionally coupling a technique known as scanning diffraction or 4D STEM\cite{ophus_four-dimensional_2019}, as well as optical properties through use of monochromation.  When applied to a random mixture of anatase and rutile nanoparticles, this approach is capable not only of mapping the spatial distribution and abundance of these polymorphs over the experimental field of view, but also directly correlating their average crystallographic information to their local chemistry and optical properties via coupled EELS and EDX datablocks.  We argue that this can be an extremely powerful technique for unravelling the complex relationship between nanoparticle morphology and polymorph distribution in this and other systems.

\section*{Results}

\subsection*{Nanoscale properties exposed by STEM}

\begin{figure}[ht]
	\centering
	\includegraphics[width=\linewidth]{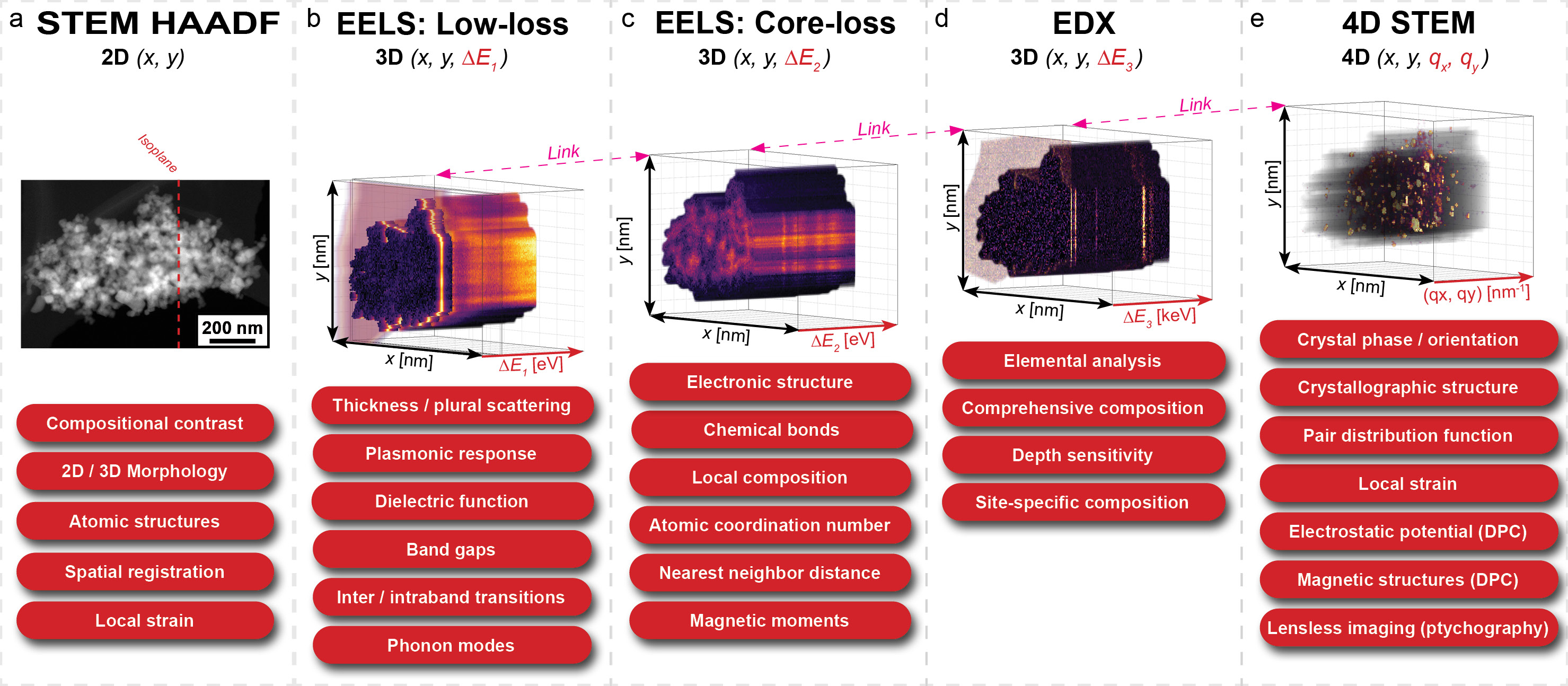}
	\caption{Schematic outlining the concept behind hypermodal data fusion. Each detector mode is labelled along with its dimensionality at top, while some of the materials properties that each technique can study are presented in red boxes below. In (a), the 2D HAADF image of the TiO$_2$ agglomerate is presented along with an isoplane that is used to expose a volumetric slice of the higher dimensional datacubes. In (b), monochromated low-loss EELS captures both the ZLP as well as the band gap region. The color scale is modified above 1 eV to account for the large dynamic range between the ZLP and inelastic scattering.  In (c), monochromated core-loss EELS captures the Ti $L_{2,3}$ edge used for polymorph feature extraction. In (d), the EDX datablock is presented, revealing compositional information about Ti, O, and impurities. Finally, (e) presents a stack of virtual dark-field images from the 4D STEM mode. The signal dimension consists of vectorized $(q_x, q_y)$ coordinates representing each pixel on the camera. Hypermodal data fusion involves extending the signal dimensions all of these datacubes (labelled in red text) by exploiting their shared spatial coordinates following a block-weighting scheme.}
	\label{fig:DFschematic}
\end{figure}

The hallmark of STEM is its ability to localize the entire electron scattering cross-section to a nano-sized volume of material. Figure \ref{fig:DFschematic} serves to illustrate which properties can be studied by exploiting these different signals, and discusses how they are specifically used in this study.

In Figure \ref{fig:DFschematic}a, a High-Angle Annular Dark Field (HAADF) image is displayed. These 2D images are generated by capturing electrons that have preferentially interacted with the nuclei of the atoms in the investigated material. Such electrons undergo elastic incoherent scattering (Rutherford scattering) where the angular dependence of the scattering cross section is roughly proportional to the square of the atomic number of the atoms \cite{pennycook_scanning_2011}. This imaging mode can thus be interpreted as a mass-thickness contrast, revealing the projected morphology of the sample. In this case, we observe a clear clustering of "heavy" nanosized particles, roughly spherical in shape, suspended upon a "lighter" platform in the background, which appears with a much weaker contrast. This is consistent with the knowledge that TiO$_2$ particles have been dispersed on a carbon supporting film, allowing for an immediate initial interpretation of the scene.

Since the HAADF image only collects high scattering angles, electrons that scatter through lower angles can be detected with additional instrumentation nearly simultaneously at each pixel position. In this case, we use an energy spectrometer to disperse the lower angle electrons according to their energy loss, a technique broadly known as Electron Energy-Loss Spectroscopy (EELS). In STEM, such datasets can be recorded with three dimensions, where the probe position is recorded as a 2D grid and the energy-loss dimension is recorded with an EELS spectrometer. Such datasets are generally known as hyperspectral datasets or Electron Spectroscopy Images (ESI). The complex energy-dependent response of the material to the introduction of a strong electromagnetic perturbation $\epsilon(E)$ can be studied using EELS \cite{egerton_electron_2011,brydson_electron_2001} as, for low energy losses, the spectrum is proportional to $\Im{(-1 / \epsilon(E))} = \epsilon_2$, permitting the investigation of the dielectric function. While this is commonly used to spatially localize plasmonic behavior, the use of monochromation can even detect optical transitions up to a few eV. The low-loss EELS (LL EELS) data from this experiment is schematically presented in Figure \ref{fig:DFschematic}b, and reveals the presence of the zero-loss peak (ZLP) as well as transitions from the valence band to the conduction band.  While this is commonly used to spatially localize plasmonic behavior, the use of monochromation permits detection of optical transitions up to a few eV.  This has been used to estimate optical band gaps in a wide variety of materials to projected morphological features on the order of a few nanometers in size\cite{reyes-coronado_phase-pure_2008,brandt_favoring_2019,zhan_nanoscale_2017,granerod_automated_2018,erni_valence_2005}, which is among the most critical properties to control in the design of nanostructured photocatalyic materials.

As the energy loss of inelastically scattered electrons increases, the imaginary part of the dielectric function $\epsilon_2$ diminishes at the expense of the real part $\epsilon_1$. EELS measured at these higher energy losses is known as core-loss EELS (CL EELS) and is similar to x-ray absorption spectroscopy techniques in that it can probe local bonding effects via Electron Energy-Loss Near Edge Fine Structure (ELNES). This effectively maps out the symmetry-projected density of unoccupied states for a given element\cite{brydson_electron_2001}. The monochromated core-loss EELS spectrum from Ti reveals the probability of electronic transitions from Ti 2p$_{1/2}$ and 2p$_{3/2}$ states to unoccupied 3d orbitals.  Critically, for titania polymorphs, the probabilities of these transitions depend on local distortions to the oxygen octahedron surrounding the metal Ti atom. This distortion reduces the point-group symmetry of the Ti atom from a perfect O$_{\rm{h}}$ to D$_{\rm{2d}}$ (anatase) or D$_{\rm{2h}}$ (rutile) \cite{brydson_electron_1989}. In both cases, this manifests itself experimentally as a splitting of the Ti $L_3$ edge into two distinct peaks ($t_g$ and $e_g$) spaced approximately 3.6 eV apart. The $e_g$ manifold, however, is particularly useful as it further splits into a doublet, hypothesized to arise from Jahn-Teller effect \cite{van_vleck_jahnteller_1939}. The spacing between these peaks is less than 1 eV, making them difficult to resolve experimentally with EELS without the use of a monochromator. However, studies on rutile and anatase have conclusively demonstrated that this feature can be used as a "fingerprint" to distinguish between the different polymorphs \cite{brydson_electron_1989} .

While monochromated EELS is best used as a targeted approach to study the localized optical response and bonding of a material, a broader picture of its elemental composition can be provided by capturing the emission of photons following decay from an excited state, which is a technique known as Energy-Dispersive X-Ray Specroscopy (EDX). In many modern TEM instruments, it is possible to collect a 3D EDX ESI simultaneously with EELS, thereby localizing these signals to the same 2D spatial registration grid (which can incidentally also be simultaneously recorded as an HAADF image). The EDX datablock in figure \ref{fig:DFschematic}e is not limited to only capturing the Ti edge, as with monochromated EELS, but also reveals the presence of C and O, trace impurities of Si and F, and even Cu.  C, Si and F are interpreted here as inconsequential contaminants present in the microscope, while Cu arises from the use of a Cu grid.  EDX is well established as a technique providing a solid assessment of the local chemical composition and can be used here to verify the Ti cross sections recorded with EELS, complement this with the O edge that was unable to be captured with this EELS configuration, check for impurities, and monitor the build-up of contamination.

Finally, while the above methods provide detailed insight into the localized chemical, optical, and electronic behavior of these titania nanoparticles, they do not directly study their local structure and crystallographic orientation. However, this is crucial to understand, as it impacts the optical performance of nanostructured catalysts, particularly at interfaces. One technique that can provide this necessary insight involves collecting a full 2D electron diffraction pattern at each scan position using a strongly parallel scanning probe, a technique resulting in a 4D dataset broadly known as scanning diffraction or 4D STEM diffraction\cite{ophus_four-dimensional_2019}. Such diffraction patterns can be indexed to determine both the crystal phase and orientation.  Additionally, this data structure can be used to generate virtual dark-field images in post processing by integrating the intensity within a virtual aperture for each real-space pixel position, as demonstrated in figure \ref{fig:DFschematic}f. Since both anatase and rutile have distinct crystal structures, it should theoretically be possible to distinguish between these phases using scanning diffraction alone. However, in practice, this is not feasible for randomly oriented particles, such as in this experiment. First, many of the primary Bragg spots are very close to each other and, thus, overlap. This effect is exacerbated by the lack of a perfectly parallel electron probe, leading to the appearance of disks rather than spots in the diffraction pattern. Second, some orientations of the nanoparticles extinguish the most easily distinguishable peaks, meaning that not all particles belonging to a specific phase will be identified in this manner.

In summary, STEM is clearly capable not only of probing an enormous range of technologically relevant materials properties, but it can moreover localize these properties to lateral areas on the order of square nanometers or less in morphologically complex systems. However, each of these properties is derived from a distinct, independently recorded dataset. Consequently, while these properties can be studied individually, it is not possible to directly study their dependencies with this basic approach. In the following sections, we discuss how these datasets can be correlated to each other using a method we call hypermodal data fusion.

\subsection*{Hypermodal data fusion workflow}

\subsubsection*{Spatial registration}

\begin{figure}[ht]
	\centering
	\includegraphics[width=\linewidth]{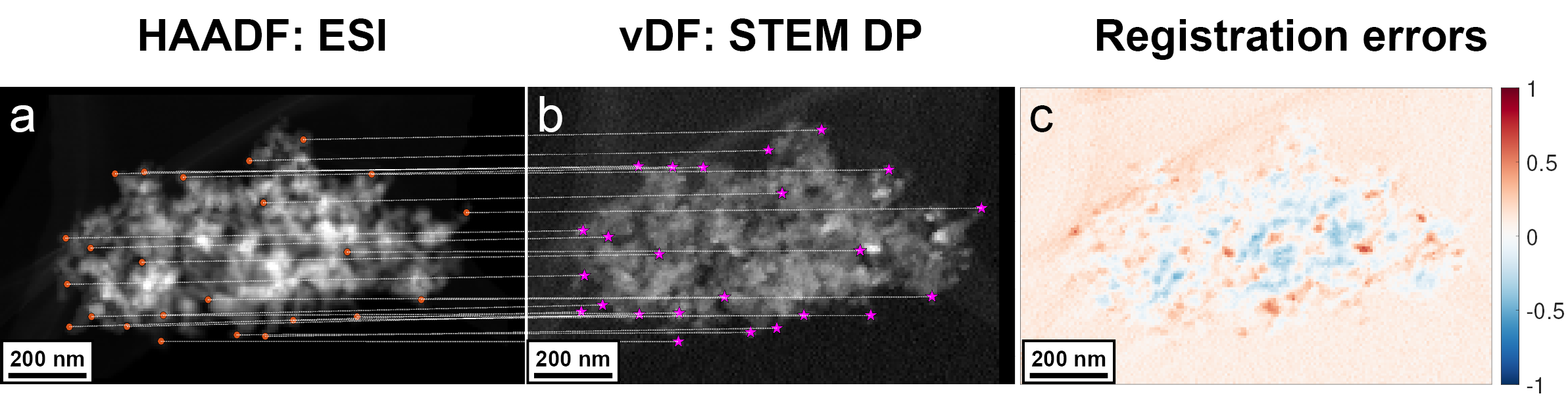}
	\caption{(a) Virtual ADF image from the scanning diffraction dataset, used as the reference image for the spatial registration. (b) HAADF image recorded while recording the EELS and EDX data, used as the moving image. (c) Difference map following registration. Both HAADF images were normalized prior to taking the difference map, so the errors can be interpreted as a pixel-wise relative error.}
	\label{fig:SpatialRegistration}
\end{figure}

A key criterion necessary to correlate these different datasets is to ensure that they are acquired from the same spatial grid.  In the case of parallel dataset acquisition, such as with all of the spectroscopy datasets, this criterion is naturally met; however, for sequential dataset acquisition, such as with scanning diffraction, the grid is distorted and/or offset but shares the same field of view and captures the same physical features.  These registration errors must first be corrected via spatial registration, which is depicted in figure \ref{fig:SpatialRegistration}.

Spatial registration for the titania dataset presented here was performed by applying a locally-weighted mean two-dimensional geometric transformation to the spectroscopy datacubes \cite{goshtasby_image_1988}.  This transformation is non-rigid and the deformation map is determined by fitting a local polynomial to 24 control points that were determined by manual inspection of morphological features common to a reference image and a moving image.  The reference image was taken to be a virtual annular dark-field map calculated from the scanning diffraction datacube (figure \ref{fig:SpatialRegistration}a), while the moving image was the High-Angle Annular Dark Field (HAADF) image recorded during the acquisition of the spectroscopy datacubes (figure \ref{fig:SpatialRegistration}b).  The resulting difference map (figure \ref{fig:SpatialRegistration}c) shows reasonably low residuals that are mostly related to intensity differences caused by using different detector modes and integrating over different angular regions.  Morphologically, the deformation map results in a registration of the spectroscopy dataset to the scanning diffraction dataset where registration errors are much smaller than the recovered features.

We note here that the spatial registration approach detailed here will not result in a "pixel perfect" alignment between datasets. However, these datasets are all oversampled with respect to the features under investigation, namely the titania nanoparticles. Accordingly, EELS spectra and diffraction patterns from within any individual nanoparticle are very similar, and the registration errors we observe are smaller than these morphological features. Thus, this step is appropriate for the relatively wide field-of-view used in this experiment, but could be expected to be problematic when moving to higher resolution or undersampled datasets.

\subsubsection*{Hypermodal data fusion}

As discussed in the introduction, hypermodal data fusion describes a methodology by which an arbitrary number of independent datasets can be combined provided they share a common spatial registration grid. While data fusion in the TEM literature has been explored primarily for spectroscopy datasets \cite{spiegelberg_signal_2018, braidy_unmixing_2019,muto_application_2019,thersleff_dissecting_2020,thersleff_soot_2021,merkl_plasmonic_2021,schwartz_recovering_2021}, we show here how this can be generalized to include any type of detector mode by including scanning diffraction. The hyperspectral datasets investigated here each consist of two real-space dimensions of length $x$ and $y$ that is common to each datablock (either naturally due to simultaneous acquisition or approximated from the spatial registration described above). Each datablock additionally consists of an arbitrary number of signal dimensions that contain the experimental results sampled at each probe position. Each individual dataset is then unfolded into a two-dimensional matrix $\mathbf{Y}_{m} \in \mathbb{R}^{s_m \times p}$ where $p = x \times y$ and $s_m$ denotes the number of data points resulting from vectorization of the signal dimension in detector mode $m$. In this manner, the columns of $\mathbf{Y}_{m}$ contain the one-dimensional signal (either spectra or diffraction patterns in this case) while the rows contain the spatial coordinates.

Following this unfolding to 2D matrices, the datasets are linked via a low-level data fusion approach via matrix concatenation \cite{van_mechelen_generic_2010,cocchi_data_2019,thersleff_dissecting_2020}, as per equation \ref{eq:MatConcat}
\begin{equation}
    \mathbf{F} = \begin{bmatrix} \frac{w_1}{\theta_1} \mathbf{Y}_{1} \\
    \vdots \\
    \frac{w_m}{\theta_m} \mathbf{Y}_{m}\end{bmatrix}
    \label{eq:MatConcat}
\end{equation}
Here, $\theta_m$ is a scalar chosen either to normalize the variance of $\mathbf{Y}_{m}$ up to a desired rank or to its total variance, while $w_m$ is an arbitrary weight applied to modality $m$.  $\mathbf{F}$ is now a large two-dimensional matrix of size $s \times p$ where $s = s_1 + \cdots + s_m$. $\mathbf{F}$ relates all of the experimental results collected from the hypermodal scene (expressed along the columns) to the spatial coordinates (expressed along the rows). It can be interpreted as a stack of vectorized energy-filtered 2D images (spectroscopy datasets) from each energy channel or virtual dark-field images (4D STEM) from each detector pixel (see figure \ref{fig:DFschematic} for a visualization).

As discussed in Thersleff et al. \cite{thersleff_dissecting_2020}, it is possible to design block weights $w_m$ such that the variance any desired detector modality can dominate the variance of $\mathbf{F}$. When subjected to dimensionality reduction, individual datablocks with the highest $w_m$ will then contribute the most to the resulting model. Since all datablocks are now linked, features extracted from individual datablocks with lower values of $w_m$ can then be interpreted as spatially correlated to the dominant datablocks. One can thus design an experiment such that the features of one dataset can be used to extract spatially correlated features in the others.
 
As the goal of this experiment is to distinguish between overlapping anatase and rutile nanoparticles, we exploit the knowledge that these polymorphs are most readily distinguished via ELNES variations in their monochromated core-loss EEL spectra. To a first approximation, the anticipated chemical rank of such a hyperspectral scene is simply two, with single-scattering EEL spectra belonging to either anatase or rutile. Accordingly, the weight for the core-loss EELS dataset is set to 1, while the weights for all other datasets were set to $1\times10^{-3}$. Prior to normalization, plural scattering was removed from the core-loss dataset using the Fourier Ratio method \cite{egerton_coupling_1976}, the pre-edge background was subtracted using a power-law model \cite{egerton_electron_2011}, and the variance was stabilized using the Generalized Anscombe Transformation (GAT) \cite{starck_image_1998,makitalo_optimal_2013}.  All other datasets were unmodified aside from normalization by $\frac{w_m}{\theta_m}$ and the standard pre-treatment procedures described in the methods section.

\subsubsection*{Singular Value Decomposition}

\begin{figure}[ht]
	\centering
	\includegraphics[width=\linewidth]{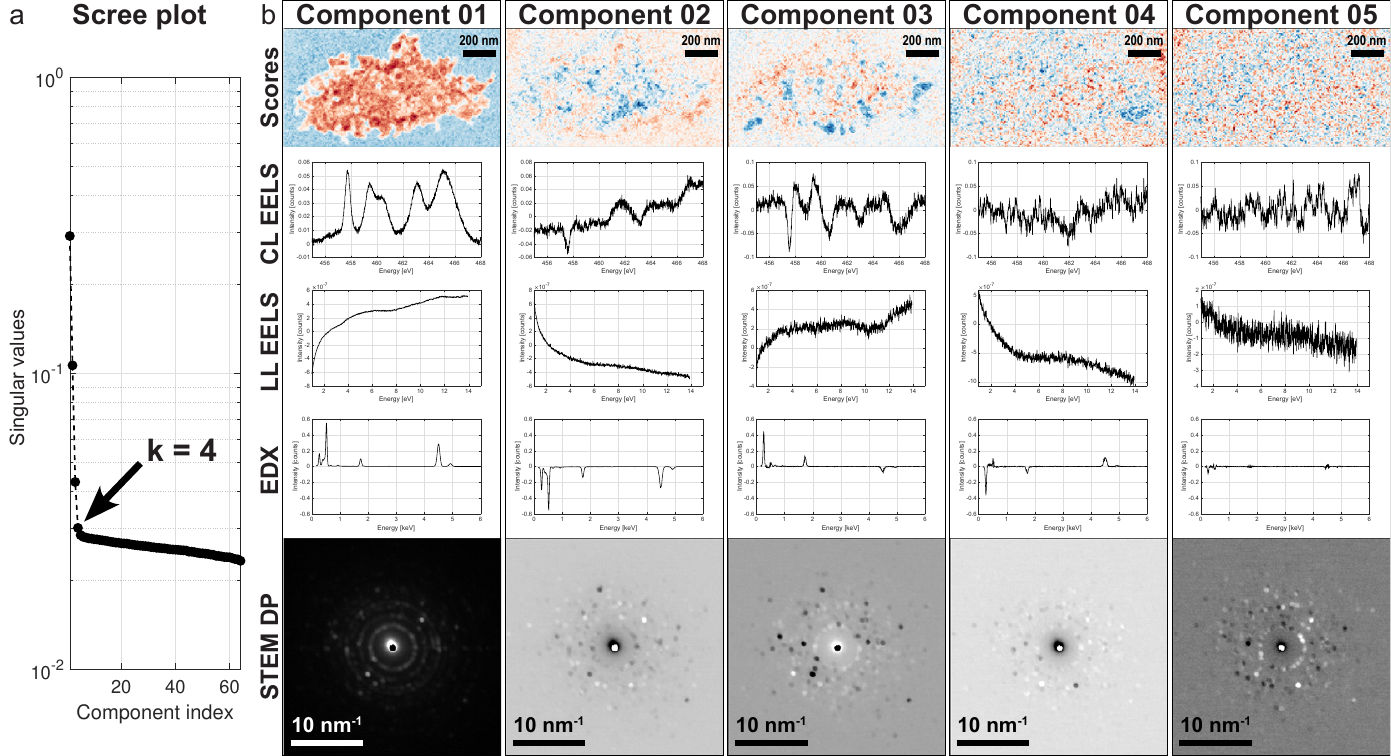}
	\caption{(a) Scree plot of the singular values from the rSVD decomposition. The estimated mathematical rank $k = 4$ is denoted. (b) Summary of the first five rSVD components using core-loss EELS results as the predictor. The score maps are displayed using a divergent color map (blue = negative, red = positive) to emphasize the sign. Core-loss EELS, low-loss EELS, EDX, and STEM DP patterns for each components are also displayed.}
	\label{fig:SVDresults}
\end{figure}

A 64-rank ($k = 64$) approximation of $\mathbf{F}$ was computed using the randomized singular value decomposition (rSVD) \cite{yu_efficient_2018,brunton_data-driven_2019} according to equation \ref{eq:SVD}
\begin{equation}
    \mathbf{F} \approx \mathbf{U}_k \Sigma_k \mathbf{V}^T_k
    \label{eq:SVD}
\end{equation}
Here, $\mathbf{U}_k$ is a unitary matrix of dimensions $s \times k$  containing the experimental results along its columns, $\mathbf{V}_k$ is a unitary matrix of dimensions ${p \times k}$ containing the spatial information along columns, $\Sigma_k$ is a non-negative diagonal matrix of dimensions $k \times k$ containing the singular values, and $\left [ \cdot  \right ]^T$ denotes the matrix transpose. Plotting the diagonal of $\Sigma_k$ vs. $k_i$ is known as the scree plot, and this provides a sense of the statistical significance for each component $k_i$. This is presented in figure \ref{fig:SVDresults}a, revealing that a reasonable approximation of the signal subspace describing $\mathbf{F}$ is achieved using the top ranking 4 components. It is important to emphasize that these 4 components only describe variations present within the core-loss EELS datablock due to the weighting scheme described above. Variations present in the other datablocks that are not spatially correlated to these ELNES variations (so-called "distinct" components) will not be retrieved in this manner, as their variance has deliberately been scaled below the noise components from the core-loss EELS dataset. If it is desired to study them, a mixed weighting scheme should be employed, as discussed in Thersleff \textit{et al.} \cite{thersleff_dissecting_2020}. 

The SVD decomposition presented in figure \ref{fig:SVDresults} provides two crucial results. First, the scree plot as well as visual inspection of all of the components provide an assessment for the mathematical rank of this system, which we estimate to be four. Second, it reduces the dimensionality of $\mathbf{F}$ to a set of four orthogonal spatial and signal latent vectors. The columns of $\mathbf{U}_k$ define the most efficient rank-4 orthogonal transformation necessary to approximate the raw data in a least-squares sense, while the columns of $\mathbf{V}_{k}$ contain the corresponding weights, also known as scores. Critically, these four components directly correlate the spatial vector (containing real-space coordinates) to the signal vector, which contains the data that will be interpreted as physical properties. Due to data fusion, this now reveals the most statistically significant spatial correlations between each individual dataset.

The scores and components for the rSVD decomposition are presented in \ref{fig:SVDresults}b. The first four score maps show significant spatial correlations, consistent with their higher singular values in figure \ref{fig:SVDresults}a. These decrease in significance until the correlations observed in component 5 are largely stochastic and convey little morphologically useful information. The loading curves for the core-loss EELS (row 2) expose significant spectral features in the first three components.  Component 01 describes the deviation from the mean, establishing the baseline, component 02 describes a change in slope that is largely localized to the thickest regions of the sample, component 03 describes shifts observed in the ELNES, component 04 captures some weak residual effects, and components 05 and above mostly describe noise. The loading curves for the LL EELS, EDX, and STEM DP datasets capture the variations that are spatially correlated to the CL EELS datablock. In these cases, we also observe that, for components 5 and above, the signals are largely devoid of information that can be interpreted in a meaningful manner, reflecting the uncorrelated nature of the corresponding score maps.

The rSVD model presented in figure \ref{fig:SVDresults} is a powerful tool to estimate the mathematical rank of this system, and this approach is a popular tool in the TEM literature for denoising purposes\cite{trebbia_eels_1990,bosman_mapping_2006,bosman_two-dimensional_2007,dudeck_quantitative_2012,potapov_why_2016,spiegelberg_usage_2017,potapov_enhancement_2017,potapov_optimal_2019}. However, it is notoriously difficult to interpret in a chemically-meaningful manner, as all of the components need to be added together subject to their weights in the score maps. Consequently, only in very rare cases will such a model capture the true, chemical rank of the system, which was the objective of the original experiment. Converting the mathematical model presented in figure \ref{fig:SVDresults} into a chemically-meaningful "true" model requires an additional processing step. 

\subsubsection*{Hyperspectral geometric unmixing}

To convert from a mathematical to a chemically meaningful model, we express $\mathbf{F}$ as 
\begin{equation}
    \mathbf{F} = \mathbf{A}_q\mathbf{S}_q + \mathbf{N}
    \label{eq:UnmixEquation}
\end{equation}
where $\mathbf{A}_q$ is a $s \times q$ matrix of pure spectral signatures, $\mathbf{S}_q$ is a $q \times p$ matrix of weights and $\mathbf{N}$ is a $s \times p$ matrix containing the additive noise component.  Here, we use $q$ to denote an estimate of the true, or chemical rank of the system (as opposed to $k$, which estimates the mathematical rank). For linear systems, equation \ref{eq:UnmixEquation} can be solved via matrix factorization, and this is often achieved through an iterative approach imposing additional constraints on $\mathbf{A}_q$ and/or $\mathbf{S}_q$, such as non-negativity to better reflect the physics of the system. However, such an approach is computationally very expensive, extremely sensitive to the starting conditions, and may not readily converge to a global minimum that can be interpreted in a physically-meaningful manner. Since $q$ represents the chemical rather than mathematical rank, the columns of $\mathbf{A}_q$ are referred to as endmembers rather than components, while the rows of $\mathbf{S}_q$ are called abundances rather than scores. This formulation of the hyperspectral scene is considerably more intuitive than its mathematical decomposition, as the endmembers can be interpreted as representative of the true nature of the experiment (provided non-linear effects such as plural scattering for EELS have been removed). 

In this work, we estimate $\mathbf{A}_q$ and $\mathbf{S}_q$ by exploiting the data geometry of $\mathbf{F}$, described by the factor space recovered in equation \ref{eq:SVD}. This approach is broadly known as geometric unmixing, and has been successfully applied to hyperspectral STEM datasets in the past \cite{dobigeon_spectral_2012,shiga_sparse_2016,spiegelberg_analysis_2017,thersleff_dissecting_2020,potapov_extraction_2021,thersleff_soot_2021}. The geometric unmixing algorithm we employ here is called Nonnegative Matrix Factorization from Quadradic Minimum Volume (NMF-QMV) \cite{tsung-han_chan_convex_2008,bioucas-dias_hyperspectral_2012,li_robust_2016,zhuang_regularization_2019}. NMF-QMV tackles the challenge of estimating $\mathbf{A}_q$ and $\mathbf{S}_q$ via an intermediate step of computing a $(q\text{-}1) \times q$ unmixing matrix $\mathbf{M}$, which combines a rotation with a projection of the latent factors from equation \ref{eq:SVD} into a $(q\text{-}1)$ affine space. It is estimated together with $\mathbf{S}_q$ via convex optimization, as shown in equation \ref{eq:UnmixEquation}
\begin{equation}
     \underset{\mathbf{M},\mathbf{S}_q}{\mathrm{min}}\frac{1}{2}\left\| \mathbf{V}_{(q \text{-} 1)}-\mathbf{M}\mathbf{S}_q\right\|^2_F + \frac{\beta}{2}\left\| \mathbf{MB} - \mathbf{O}\right\|^2_F
    \label{eq:NMFQMV}
\end{equation}
In this notation, $\mathbf{V}_{(q\text{-}1)}$ is the affine projection of the score matrix $\mathbf{V}$ from equation \ref{eq:SVD} into a $(q\text{-}1)$ subspace, while $\mathbf{B}$ and $\mathbf{O}$ are boundary conditions describing the volume of the geometric simplex defined by the verticies of $\mathbf{M}$ contained along its columns (details can be found in Zhuang et al. \cite{zhuang_regularization_2019}). Equation \ref{eq:NMFQMV} balances the minimization of data reconstruction erorrs (the first term) with the minimization of the simplex volume (the second term) through the introduction of the regularization parameter $\beta$. Larger values of $\beta$ favor volume minimization over reconstruction errors and, thus, can be used for targeted outlier rejection. $\mathbf{A}_q$ is subsequently recovered via equation \ref{eq:UnmixingMatrix}
\begin{equation}
    \mathbf{A}_q = \mathbf{U}_{(q \text{-} 1)}\mathbf{M}
    \label{eq:UnmixingMatrix}
\end{equation}
During this optimization, additional constraints can be applied to $\mathbf{S}_{q}$, such as non-negativity and sum-to-one ($\mathbf{1}^T_q\mathbf{S} = \mathbf{1}^T_p$ where $\mathbf{1}^T$ is a column vector of ones having span $p$).

Formulating the factorization in this manner is hugely beneficial for both computational and interpretive purposes. Computationally, since $\mathbf{M}$ is calculated in the factor space estimated from the dimensionality-reduced rSVD model, thousands of iterations can be rapidly performed, even if $\mathbf{F}$ is many tens of gigabytes in size. Combined with use of rSVD, this means that this entire workflow can be completed in a time scale of seconds to minutes on an average desktop computer (provided sufficient physical ram is installed in the system to load $\mathbf{F}$). Conceptually, since the vertices of $\mathbf{M}$ describe a geometric object in the factor space of $\mathbf{F}$, it can provide immediate and intuitive visual feedback pertaining to the feasibility of the model. An analogy for a rank 3 system would be a ternary phase diagram, where any datapoint within the projected 2D ternary simplex (a triangle) is a weighted combination of the compounds defined by its three vertices. We explore this conceptual interpretation for the titania dataset in figure \ref{fig:UnmixingResults}.

Although most geometric unmixing algorithms use variations of the simplex volume minimization approach outlined above, NMF-QMV has one additional property that makes it particularly relevant for hypermodal STEM experiments, namely that the vertices of $\mathbf{M}$ are allowed to extrapolate outside of the data cloud. Physically, this represents the case when a given endmember is never recovered in its pure or unmixed form in the raw data (similar to never finding a pure compound in the ternary phase system analogy introduced above). While this assumption is not necessary for hypermodal STEM experiments on thin samples where the endmembers are spatially isolated\cite{spiegelberg_analysis_2017,potapov_extraction_2021}, it is a much more reasonable assumption to make for complex morphologies such as core-shell heterostructures\cite{thersleff_dissecting_2020,thersleff_soot_2021} or distinct particles that overlap each other projection in the 2D projection of the STEM experiment, where the maximum abundance of a given endmember never reaches unity. NMF-QMV accounts for this situation by fitting the simplex object to the facets of the data cloud (estimated from its convex hull) rather than placing them at estimates of its geometric extremeties (similar to clustering). The emphasis of facets rather than extremeties naturally makes NMF-QMV less sensitive to outliers which is highly advantageous for the noisy data typical for rapidly-acquired hypermodal STEM experiments, but places a heavier emphasis on proper interpretation of the resulting model, as precise placement of vertices located further away from the data cloud can be expected to be less reliable than with a clustering approach. We discuss the ramifications of this in greater detail in the discussion section.

\subsection*{Unmixing results}
\subsubsection*{Estimate of the unmixing matrix}

\begin{figure}[ht]
	\centering
	\includegraphics[width=1\linewidth]{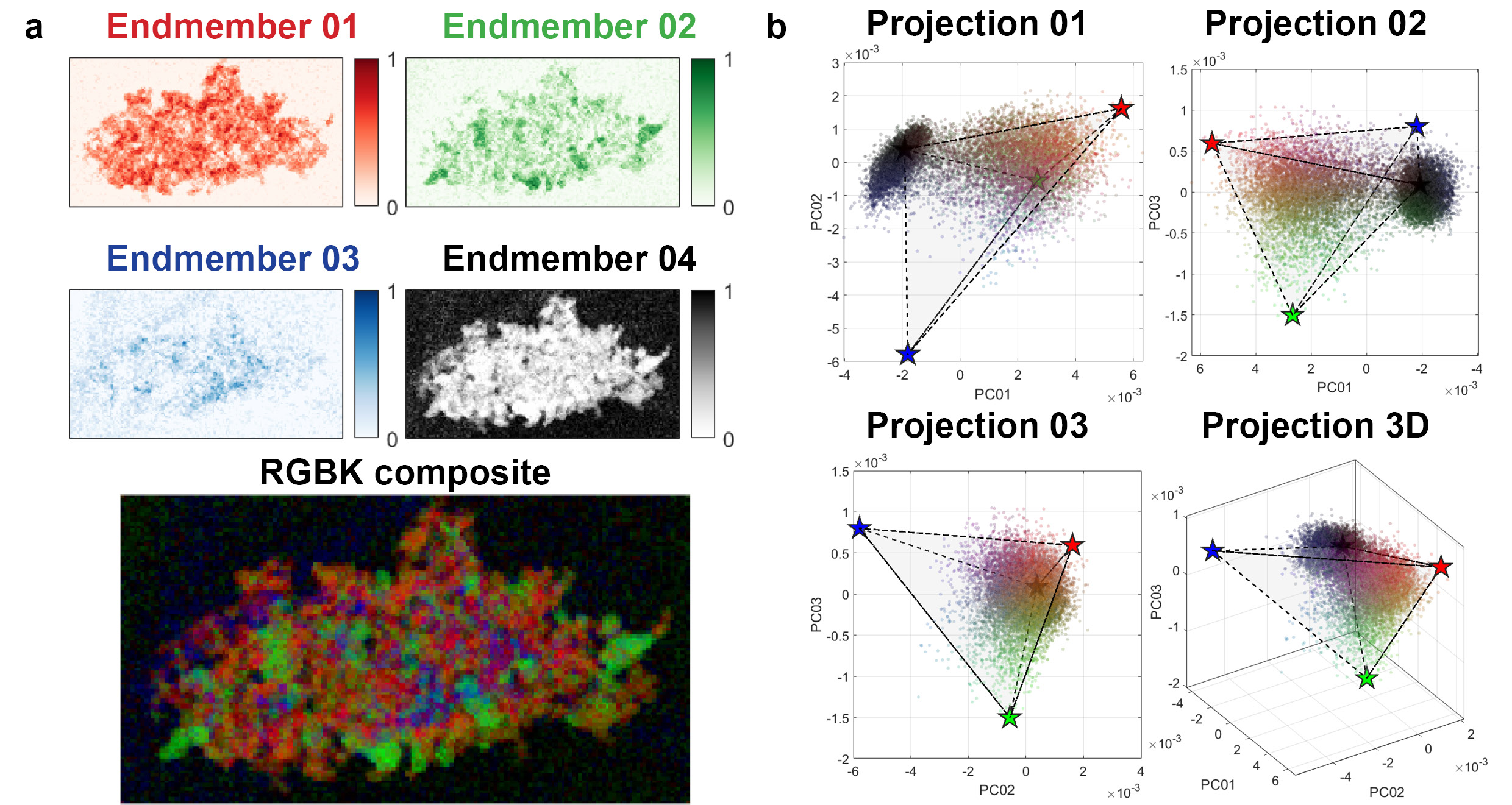}
	\caption{(a) Data geometry for the joint dataset.  The rank-four approximation used in this paper can be represented as a 3D simplex, and the three principal projections are shown here. The orthogonal axes of these graphs are represented by the principal components (PC) presented in figure \ref{fig:SVDresults}.  The unmixing matrix itself is represented as a 3D simplex in this dataset, and the vertices are labeled with the corresponding endmember index.  The endmembers themselves are presented in figure \ref{fig:UnmixingResults}.}
	\label{fig:DataGeometry}
\end{figure}

For the hypermodal titania dataset described in this paper, the unmixing matrix $\mathbf{M}$ is estimated from the core-loss EELS datablock, as the other datablocks have been weighted such that their variance does not contribute to the rSVD model in figure \ref{fig:SVDresults}. By manually trying many values of $q$, we find that the most parsimonious estimate for the chemical rank of this system is $q = 4$.  It is worth noting that, while the estimate of $q$ can be automated by promoting sparsity along the rows of $\mathbf{S}_{q}$ \cite{li_robust_2016}, our decision here is based off of our scientific experience interpreting the explanatory power of the retrieved endmembers. We motivate this strategy further in the discussion section.

The data geometry is presented in figure \ref{fig:DataGeometry}b as a series of scatter plots of $\mathbf{V}_{(q\text{-}1)}$ in the orthogonal basis described by the columns of $\mathbf{U}_{(q\text{-}1)}$ (labelled "PC" here for principal component). The data points are colored according to the derived abundance maps, which are provided in the RGBK composite image in figure \ref{fig:DataGeometry}a. The 3-dimensional simplex described by $\mathbf{M}$ is overlaid on these data and its vertices are also colored according to the endmember they describe.

Figure \ref{fig:DataGeometry} allows us to make a few key observations.  First, the data appear to cluster into two distinct data clouds.  The densest of these clouds is colored more dark and it corresponds to the Carbon support film and vacuum in hyperspectral scene (endmember 04).  The spread of datapoints in this cluster visually depict the magnitude of the noise present in this region of the hyperspectral dataset. Since the SNR in this region is higher than on the titania agglomerate due to the overall lack of core-loss EELS counts on the titanium edge, NMF-QMV originally estimates the vertex position to be further away from this data cloud. This leads to negative values in the endmembers and is difficult to interpret physically. Consequently, we relax the constraint on reconstruction errors for this vertex. In other words, we allow this vertex to be defined not by the facets of the data structure but, rather, by a weighted average of this cluster, suggesting that the vertex belonging to endmember 04 should be placed directly within the cloud. This makes sense physically, as this feature has an abundance of nearly 1 when the titania particles are spatially absent, but means that this approach can no longer be considered fully unsupervised. We explore this topic more in the discussion section below.

The second, larger data cloud contains the information needed to describe the chemistry of the titania nanoparticles, which is the objective of this experiment. These data points have a more clear geometry associated with them, with reasonably well-defined facets.  The simplex defined by $\mathbf{M}$ is estimated by following these facets and extrapolating outside of the data. The number of datapoints outside of this volume is set by adjusting $\beta$ from equation \ref{eq:NMFQMV} and was optimized by manual inspection of the resulting endmembers for conditions that lead to the most self-consistent physically-meaningful model for all data blocks (not just core-loss EELS). We observe that the vertex positions for endmembers 01 and 02 are highly consistent over multiple iterations, probably due to their proximity to the data cloud, giving us high confidence in their placement. Endmember 03, however, was more challenging due to its placement further away from the data cloud. Nevertheless, we were unable to find a reasonable model for the data where this vertex was excluded, so we include it here along with our interpretation.

\subsubsection*{Datablock-specific endmember estimation}

\begin{figure}[ht]
	\centering
	\includegraphics[width=\linewidth]{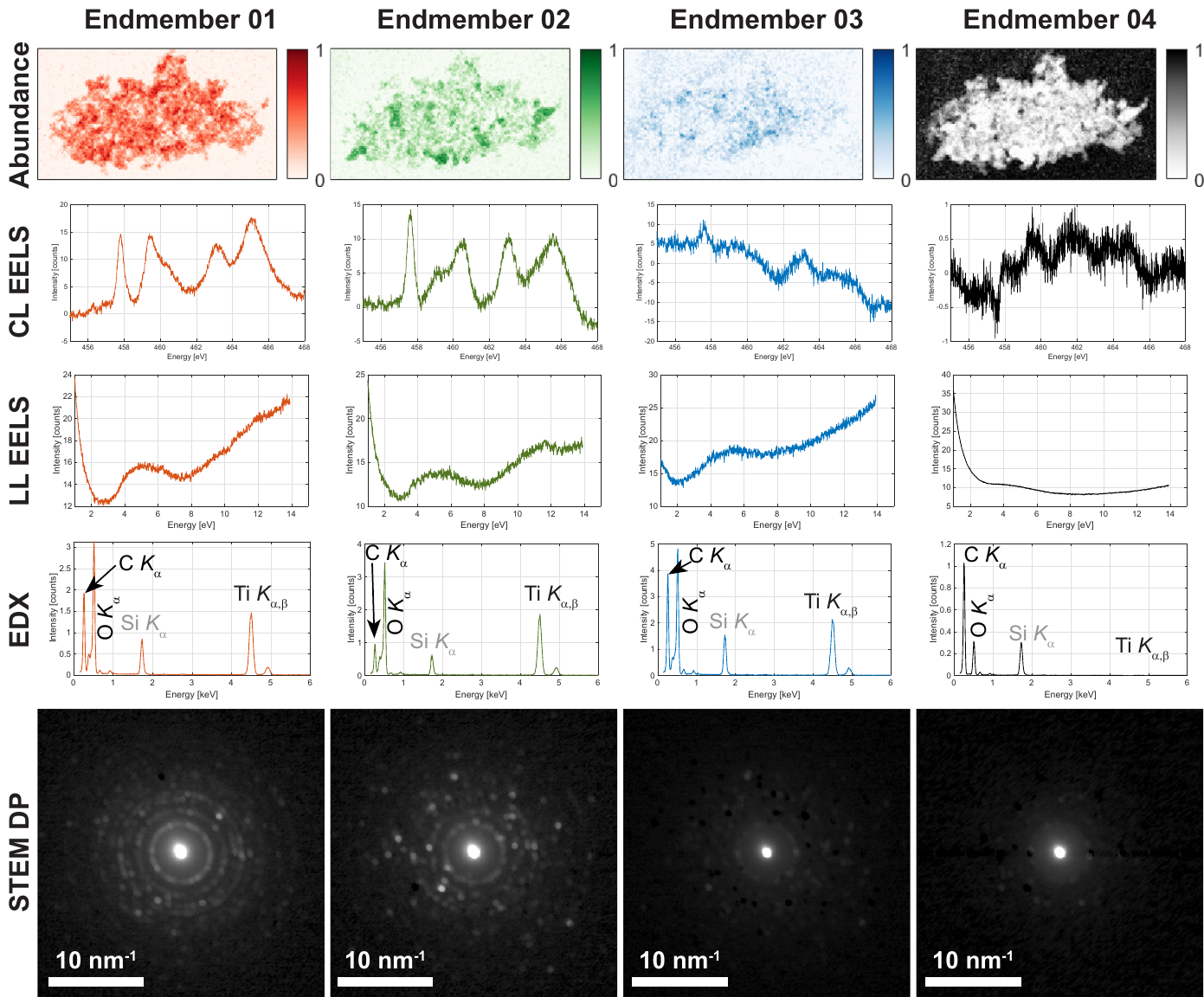}
	\caption{Summary of the results from geometric unmixing}
	\label{fig:UnmixingResults}
\end{figure}

\begin{table}[ht]
\centering
\caption{Summary of the endmember abundances.}
\label{tab:Abundance}
\begin{tabular}{lllll}
\hline
Endmembers & 01   & 02   & 03   & 04   \\ \hline
Sum        & $2.95 \times 10^3$ & $2.22 \times 10^3$ & $1.21 \times 10^3$ & $8.89 \times 10^3$ \\ 
Percent    & 19.3\% & 14.5\% & 7.9\% & 58.2\%  \\ \hline
\end{tabular}
\end{table}

A summary of all the unmixing results for all of the fused data blocks is presented in figure \ref{fig:UnmixingResults}. The abundance maps shown in figure \ref{fig:DataGeometry}a are reproduced in the first row here for clarity. Since these were subjected to nonnegativity and sum-to-one constraints, they are scaled here from 0 to 1. This also allows for the relative abundance of each endmember to be estimated, and this is provided in table \ref{tab:Abundance}. We observe that the ratio of endmembers 01 and 02 is approximately 0.57, which represents the approximate ratio of anatase to rutile in this system (0.67).  A difference in morphologies between endmembers 01 and 02 is also observed. The abundance distribution for endmember 01 is homogeneously distributed over the entire agglomerate and appears to consist of a large number of smaller particles. In contrast, the abundance distribution for endmember 02 is highly concentrated in a number of regions which can be morphologically related to individual particles, as observed in the HAADF image presented in figure \ref{fig:DFschematic}a. A more detailed analysis of the different signal domains follows.

\subsubsection*{Core-loss EELS}
The core-loss EELS endmembers associated with their respective abundance maps represent the most parsimonious model required to approximate the raw core-loss EELS data with only four endmembers while obeying the aforementioned constraints and respecting the data geometry presented in figure \ref{fig:DataGeometry}.  Using the ELNES fingerprinting approach, we observe that the retrieved core-loss EELS spectrum from endmember 01 bears a striking resemblance to reference spectra acquired on pure anatase while endmember 02 closely resembles rutile\cite{brydson_electron_1989}. This strongly suggests that NMF-QMV can use the core-loss EELS as a way to estimate the unmixing matrix necessary to isolate these chemical features in factor space.

Endmembers 03 and 04 capture the distribution of spectra that are largely unassociated with the Ti $L_{2,3}$ edge.  Endmember 04 is fairly straightforward to interpret, as it largely describes the variations that arise from the background. Due to the deliberate placement of the vertex within the dense data cloud, this effectively represents a weighted average of the background, similar to what would be expected by masking away the agglomerate and averaging out the ESI. A slight increase in the core-loss EELS signal at the Ti $L_3$ onset can be explained as an afterglow effect on the spectrometer, caused by rapidly moving between regions of high (Ti particles) and low (Carbon layer / vacuum) scattering potential for the recorded energy range. Additional, higher frequency spectral variations are also visible, and these appear to be largely associated with residual correlated noise present in the scintillator. In this manner, the inclusion of this endmember in the overall model can help to remove some of the camera artefacts that would otherwise contribute to the other core-loss signals.

Endmember 03 is the most complex to interpret physically. First, we see that the EELS signal shows a slight decay at higher energy-loss, although some small variations are observed that may be associated with ELNES fluctuations.  Its abundance map is also roughly positively correlated to increases in thickness for the agglomerate. We therefore believe that this is likely related to the plural scattering contribution. However, this was surprising, as we had attempted to remove plural scattering in the core-loss EELS via Fourier Ratio deconvolution. Our favored physical interpretation at this stage is that this endmember represents errors caused by a combination of inaccuracies in this deconvolution and uncertainty in placement of the vertex describing this endmember. Having said this, variations in the placement of this vertex do not appear to alter the overall interpretation of the other endmembers and, critically, do not impact the physical conclusions of this manuscript. Thus, we include it here for discussion and transparency.

\subsubsection*{Low-loss and valence EELS results}

\begin{figure}[ht]
	\centering
	\includegraphics[width=\linewidth]{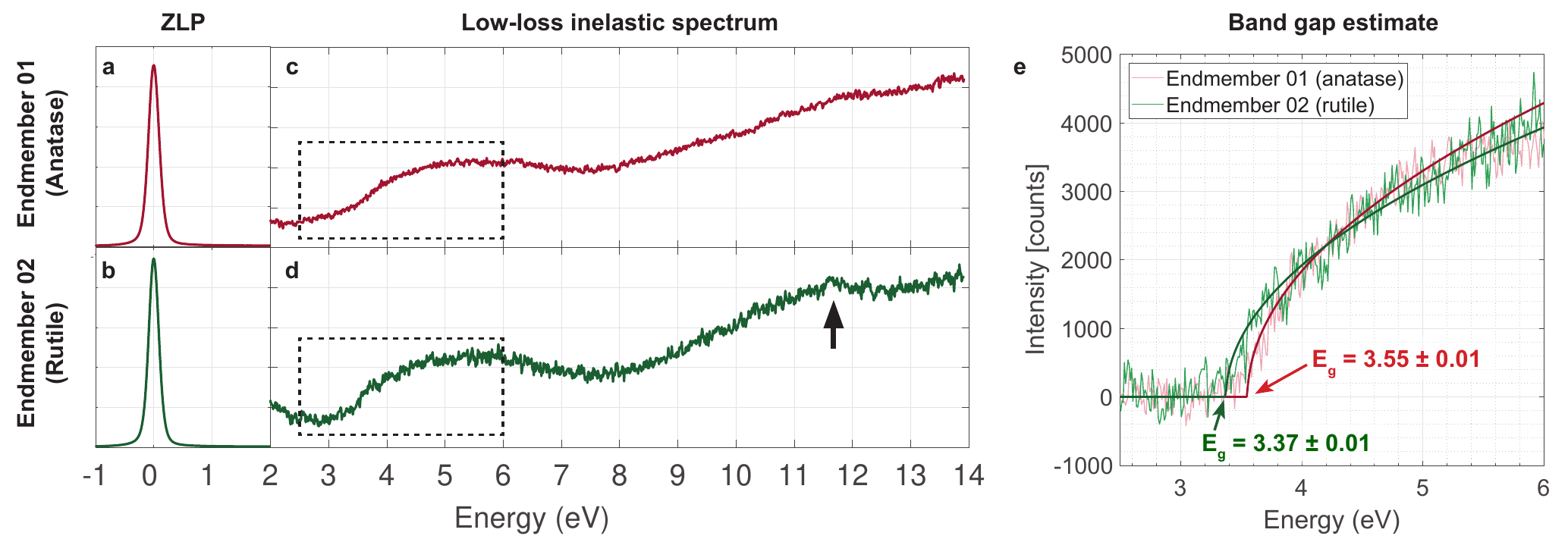}
	\caption{ZLP for endmember 01 (a) and 02 (b). The corresponding low-loss EELS spectrum from endmember 01 is presented in (c) and endmember 02 is presented in (d). An arrow denotes a spectral feature that is observed in rutile\cite{launay_evidence_2004}. The boxed region is magnified in (e) to highlight differences in the estimated bandgap from these spectra. These are estimated by a Tauc plot, which is overlaid in bold on the corresponding spectra. Errors bars represent the statistical error at 95\% confidence for the $E_g$ parameter and do not account for systematic errors.}
	\label{fig:BandGap}
\end{figure}

Although the low-loss EELS datablock was not weighted to contribute to the overall rSVD model, the exploitation of a shared spatial registration directly reveals the spatial correlations between this and the core-loss EELS datasets. Consequently, the unmixing results presented in the third row of figure \ref{fig:UnmixingResults} describe how the chemical features recovered from core-loss EELS (specifically the titania polymorphs) also vary in their monochromated low-loss and valence EELS response, implying that it is possible to isolate the polymorph-specific optical response in a strongly mixed system. 

To investigate this prospect further, we present the low-loss EELS spectra from endmembers 01 and 02 in figure \ref{fig:BandGap}.  While both spectra initially look very similar, there are small yet significant differences between them. First, the spectrum from endmember 02 (figure \ref{fig:BandGap}d) shows a small peak located at approximately 11.5 eV. This peak, denoted by a small arrow, is notably absent from the anatase spectrum in figure \ref{fig:BandGap}c. Such behavior has been observed in rutile, where it was associated with Ti$_3$O$_{15}$ clusters, and has even been proposed as an alternative fingerprinting approach to separate anatase and rutile\cite{launay_evidence_2004}.  

The ability to separate these VEEL spectra in a mixed system also implies that it might be able to estimate feature-specific bandgaps. This has major implications for investigating bandgap variations in multiphase systems, which at present are limited to measuring bandgap variations over the entire field of view despite the prospect of signal mixing \cite{zhan_band_2018}. The presence of different phases is therefore often inferred on the basis of observed bandgap variations, but it is challenging to estimate a true bandgap value in complex, overlapping morphologies. 

Prospects for measuring polymorph-specific bandgaps in titania are explored further in figure \ref{fig:BandGap}e. Here, the ZLP has been removed by fitting a logarithmic function in the Digital Micrograph software package. This graph shows both the recovered VEELS spectra for anatase and rutile in the energy range 2.5 - 6.0 eV along with Tauc fittings for both direct and indirect bandgaps according to Rafferty \cite{rafferty_direct_1998}. Here, we only report estimates of the direct band gap for both materials, which can be expected to be reasonably accurate for sufficiently thin samples\cite{gu_band-gap_2007}. The reported estimates for anatase and rutile are similar to their expected values, with anatase showing a clearly larger value. This is consistent with what is known about these materials' optical properties as well as what has been observed previously in EELS experiments\cite{brandt_favoring_2019}. 

We note here that, while we clearly see distinct differences in the VEEL spectra from endmembers 01 and 02, estimating the true band gaps from VEELS data is highly complex, requiring careful experimental design and data interpretation \cite{gu_band-gap_2007,erni_valence_2005}. We discuss the feasibility for interpreting these fits as band gaps in more detail in the discussion section. 

\subsubsection*{EDX results}

\begin{figure}[ht]
	\centering
	\includegraphics[width=0.33\linewidth]{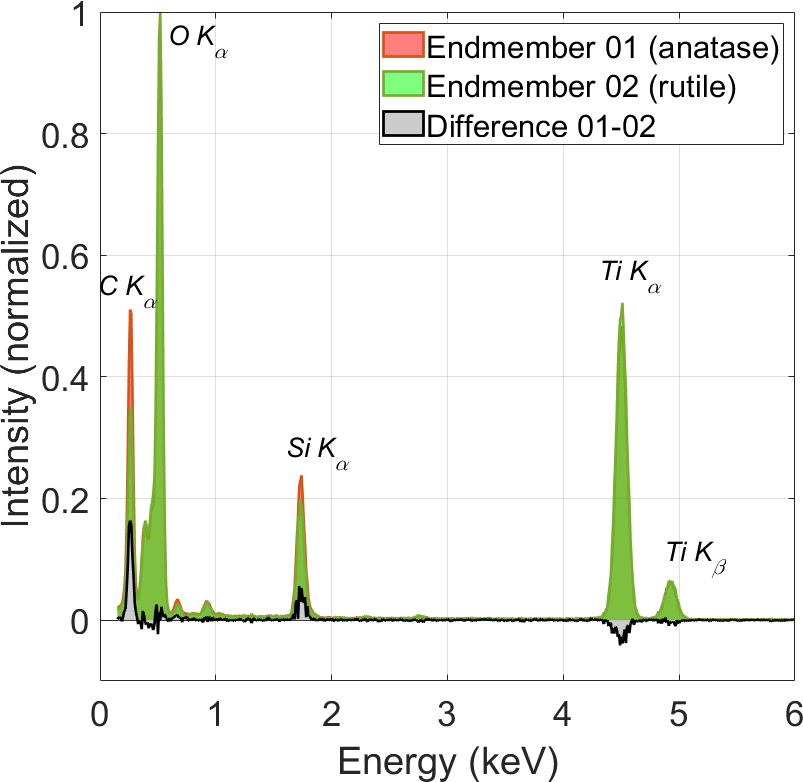}
	\caption{Comparison between the retrieved EDX spectra for Endmembers 01 (anatase) and 02 (rutile), along with their difference. Spectra have been normalized to the maximum value on the O-$K_\alpha$ edge}
	\label{fig:EDXcomparison}
\end{figure}

\begin{table}[ht]
\centering
\caption{EDX quantification results for all endmembers. Elemental results are provided in atomic percentage and are predicted to be accurate to 10\%.}
\label{tab:EDX}
\begin{tabular}{lllll}
\hline
Endmembers & 01   & 02   & 03   & 04   \\ \hline
O          & 62.0 & 61.6 & 60.7 & 51.8 \\ 
Si         & 10.8 & 8.8  & 14.4 & 48.2  \\ 
Ti         & 27.2 & 29.7 & 24.9 & 0.0 \\ 
O : Ti     & 2.3  & 2.1  & 2.4  & -  \\ \hline
\end{tabular}
\end{table}

As noted in the introduction, the core-loss EELS results only capture the Ti $L_{2,3}$ edge and are therefore not capable of estimating the oxygen composition. However, this estimate can be performed by inspecting the endmembers retrieved from the co-registered EDX dataset. Similar to the interpretation of VEELS, we now explore the prospect that the EDX spectra are also feature-specific. In the case of two titania polymorphs, we would expect this to yield a similar O : Ti of 2.

Figure \ref{fig:EDXcomparison} plots the EDX results from endmembers 02 and 04 together along with their difference following normalization to the maximum value in the O~$K_\alpha$ peak. We observe that, in both cases, similar peak intensities for Ti~$K_\alpha$ and $K_\beta$ are observed, although it is slightly lower for endmember 02 than for 01. In addition to Ti and O, we also observe a significant signal coming from Si and C, suggesting either that they are present in these phases or that $\mathbf{M}$ was unable to fully unmix them. We favor this second possibility, since variations to the Ti $L_{2,3}$ edges that were used to estimate $\mathbf{M}$ would not be expected to change appreciably in the presence of these impurities.

Quantification of the EDX results is presented in table \ref{tab:EDX}. For both endmembers 01 and 02, we estimate a O : Ti ratio that is slightly higher than 2.0. We believe that this might be due to the inclusion of residual O in the Si and C contaminants in the form of SiO$_2$ or carbonates on the particle surfaces. This may be supported by the observation that endmember 03 shows an even higher O : Ti ratio, and which is consistent with our interpretation that this endmember primarily describes a "residual" contribution from plural scattering and contamination.  A significant amount of both O and Si are moreover uncovered in endmember 04, which primarily describes the carbon support film, and they are also observed in this region in the raw data.

\subsubsection*{Scanning diffraction}

\begin{figure}[ht]
	\centering
	\includegraphics[width=\linewidth]{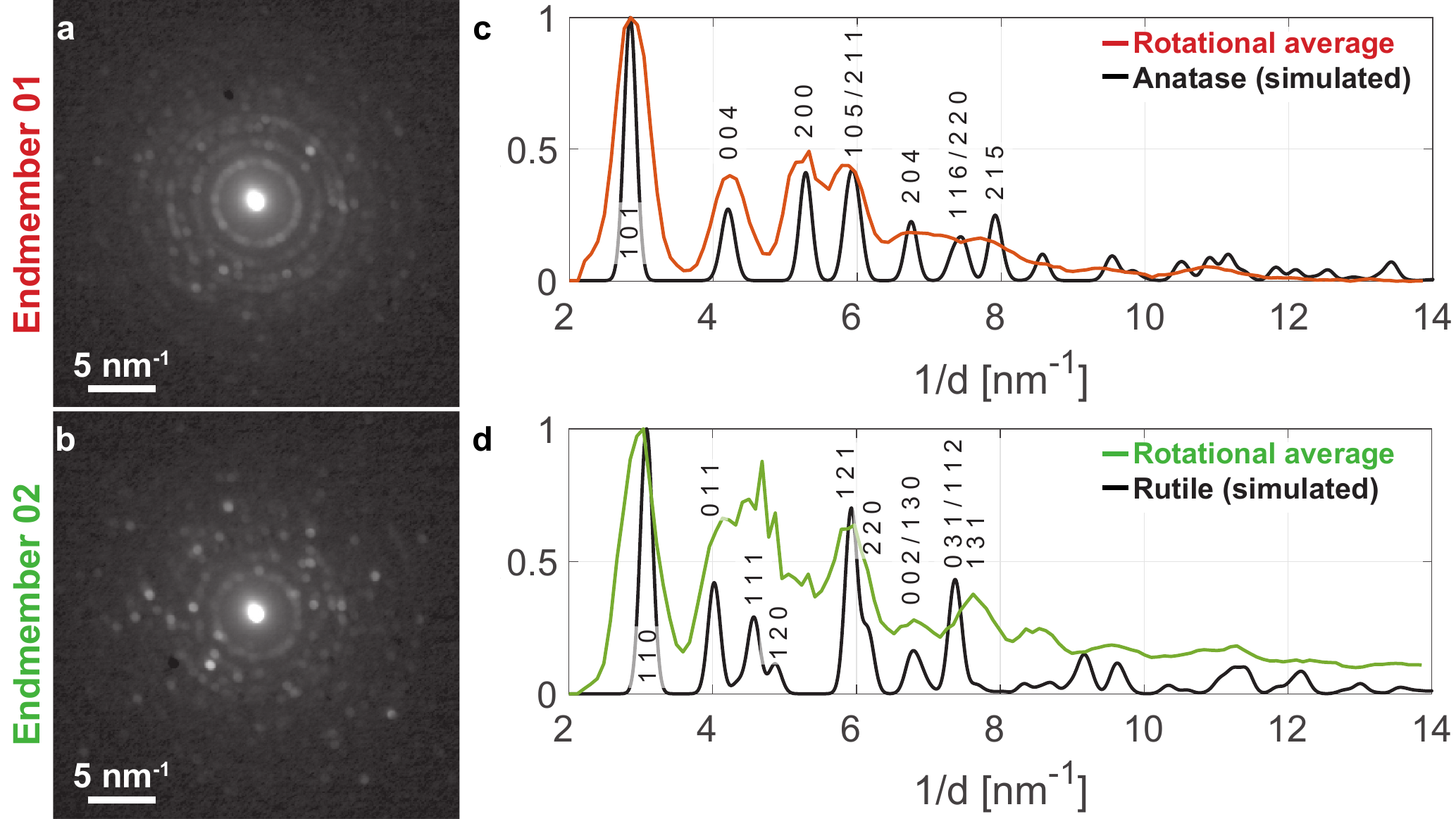}
	\caption{Unmixed 2D diffraction patterns corresponding to (a) endmember 01 and (b) endmember 02. In (c), the rotational average of (a) plotted against a simulated powder diffraction pattern for anatase, while in (d) the rotational average of (b) is plotted against a simulated powder diffraction pattern for rutile. Both (c) and (d) are normalized to their maximum intensities above $2~\textup{nm}^{-1}$ with indexing up to $8~\textup{nm}^{-1}$}
	\label{fig:RotationalAverageResults}
\end{figure}

While the previous results all describe some form of spectroscopy, and they were all acquired in parallel, the inclusion of scanning diffraction allows us to test the level of agnosticism that hypermodal data fusion can tolerate. Not only were these data acquired sequentially and with different electron optical conditions to EELS / EDX, but they also represent a much higher dimensional space that is notoriously difficult to compress. Nevertheless, the primary assumption of the hypermodal data fusion workflow - namely that the datasets describe related features from a shared spatial coordinate system - is sufficiently powerful to allow for the core-loss EELS data to retrieve the correlated mean diffraction pattern from each identified feature.

The 2D scanning diffraction patterns predicted for endmembers 01 and 02 are presented in figure \ref{fig:RotationalAverageResults}a and b, respectively. In figure \ref{fig:RotationalAverageResults}a, the pattern consists of well defined rings at distinct d-spacings, indicative of a large number of randomly oriented particles. In contrast, the diffraction pattern for endmember 02 shows considerably fewer reflections. This would be expected for a smaller number of particles, and is consistent with the findings that endmember 02 only represents 34\% of the abundance within the agglomerate according to table \ref{tab:Abundance}.

To test whether these diffraction patterns are representative of anatase and rutile, we present their rotational averages along with simulated powder diffraction patterns in figure \ref{fig:RotationalAverageResults}c and d, respectively.  Qualitatively, the match between both phases is quite good. The first observed d-spacing for endmember 01 is located at a slightly larger d-value than for endmember 02. The observed reflection for endmembers 01 and 02 fit well with anatase $0~1~1$ and rutile $1~1~0$, respectively.  Higher order reflections also largely follow the expected reflections for a simulated powder diffraction pattern for anatase and rutile. Reflections observed above 8~nm$^{-1}$ are very weak. A small discrepancy between the experimental and simulated patterns appears for endmember 02 in figure \ref{fig:RotationalAverageResults}d with the observance of the $1~3~1$ reflection in the experimental data despite having low intensity in the simulation. However, such discrepancies are expected for such a small number of nanoparticles. In this case, we likely simply scanned over a single rutile nanoparticle that yields a strong $1~3~1$ diffraction spot due to texture, i.e. deviated from power diffraction case. Thus, the good agreement between simulation and experimental patterns leads us to conclude that these patterns closely resemble the average, polymorph-specific diffraction patterns from this dataset.

\section*{Discussion and Concluding Remarks}

Based on the findings of this study, we argue that the hypermodal data fusion workflow presented here has major advantages for scientists whose questions involve unmixing disparate materials properties on the nanoscale. First, we demonstrate how it is possible to combine an arbitrary number of highly disparate co-registered hypermodal STEM datasets by linking them through their shared spatial coordinates, including a strategy for spatially registering datasets acquired in series. This is a powerful demonstration of the agnosticism of this approach, thus opening the door for a much wider range of materials properties to be investigated in the future. Second, we demonstrate how data fusion can be combined with NMF-QMV to extract an intuitive model describing the hyperspectral scene that can even partially account for the projection problem inherent in all STEM experiments. By combining this with data fusion using a tailored block weighting scheme, we show how one can identify features or properties from one dataset and impose these onto the attached datasets. This allows for the design of completely novel hypermodal STEM experiments that have not previously been possible.

Beyond merely proposing this workflow, we directly demonstrate how it can be used to directly address a highly relevant materials science problem by applying it to a challenging dataset. By complementing three parallel-acquired STEM spectroscopy datasets with scanning diffraction, we can derive insight into the average crystal structure from a random assortment of mixed polymorph TiO$_2$ nanoparticles. Moreover, we explore the prospects for using this same approach to estimate polymorph-specific compositions and even band gaps in a morphologically mixed system.

Finally, we discuss how one can interpret results such as these.  We begin by investigating the CL-EELS results, as the variance of this datablock defined the data geometry and subsequent unmixing of the other datasets. Using the ELNES fingerprinting approach and referring to literature \cite{brydson_electron_1989}, we interpret endmember 01 to belong to anatase while those from endmember 02 belong to rutile.  We moreover expect the core-loss EELS spectra extracted from these features to closely resemble the polymorph-specific single-scattering distributions, as plural scattering has been removed prior to unmixing and residual errors appear to have largely been described in endmember 03. Endmember 04, as mentioned previously, describes the afterglow effect on the camera that is prominent in the carbon support film, but also captures camera artefacts, essentially "cleaning up" the other endmembers.

The results from VEELS presented in figure \ref{fig:BandGap} strongly suggest that the hypermodal data fusion approach described here could ultimately be used to study feature-specific bandgaps in highly mixed systems. While unprecedented, making this approach reliable for band gap measurements will require some additional development. Since this experiment used an acceleration voltage of 300~kV and a rather large collection angle of 23~mrad, it is expected that retardation effects caused by Cerenkov radiation will contribute some intensity to the VEEL spectra in this energy range\cite{stoger-pollach_influence_2007,stoger-pollach_optical_2008}. This complicates the interpretation of bandgaps using the Tauc fitting approach presented in figure \ref{fig:BandGap}e and invites a more cautious interpretation of their physical meaning. However, future experiments could be designed with this in mind, for example through the use of lower acceleration voltage, and none of the workflow described above precludes this.

The results from EDX presented in figure \ref{fig:EDXcomparison} and table \ref{tab:EDX} also show great promise for this technique in estimating the full composition of individual components in highly mixed systems. In this case, the detection of Si and C over the entire sample complicates the analysis, as these could not be cleanly unmixed with this approach. However, future experiments can account for this with cleaner sample preparation protocols. Moreover, it is possible that the contamination could be identified as a feature and, subsequently, extracted using a mixed weight approach\cite{thersleff_dissecting_2020}.

The extraction of polymorph-specific 2D diffraction patterns in figure \ref{fig:RotationalAverageResults} not only highlights just how general this workflow is, but also could be used to study orientation-dependent optical properties. The most significant challenge on this front is that the scanning diffraction datablock requires a high rank rSVD model to accurately capture its variations. However, for systems where the diffraction patterns can be represented by a low-rank model or where sub-block extraction can retrieve low-rank features, it should be possible to test how orientation can influence optical properties. This is crucial to understand in advanced photocatalytic heterostructures and we feel that this approach can be very valuable to the broader community.

At this stage, we feel that there are two large challenges facing the adoption of this workflow. First, it will be necessary to modify the NMF-QMV algorithm used here to account for differences in SNR for different features. This is particularly evident with endmember 04, which required manual adjustment to achieve a physically-meaningful result. This approach is both labor intensive and also subject to human bias, and therefore should be improved. The second challenge is establishing a best-principles approach for how to interpret these results in a physically-meaningful manner. While, in our case, endmembers 01, 02, and 04 are fairly straight forward in their interpretation, endmember 03 poses a much larger challenge. We believe that this endmember is required to account for plural scattering, although this was admittedly not an expected result, as we had attempted to remove this effect prior to unmixing via Fourier ratio deconvolution. However, it is worth noting that, for monochromated EELS, only a small amount of the plural scattering contribution is recorded in the low-loss EELS dataset and, thus, we can expect some errors to arise here. This leads to a less intuitive interpretation of some of the features, with the core-loss EELS representing a negative change in slope. The LL-EELS results for endmember 03 shows a feature that could be interpreted as a band gap; however, we believe that this instead represents the average spectrum from the thickest regions of the agglomerate. This is further supported by the EDX spectrum from endmember 03, which is close to the average spectrum from the agglomerate (although it has a higher O signal, perhaps due to its preference for capturing the contamination). A deeper understanding of this phenomenon should become evident following further work in this field.

In conclusion, we have proposed and demonstrated a workflow that can successfully fuse together an arbitrary number of hypermodal STEM datasets. We show how this can be used to not only extract disparate materials properties from nano-sized volumes of material, but also correlate their \textit{relationships} in a quantitative manner via a shared spatial registration grid. We demonstrate how this can be used to solve a materials challenge by unmixing anatase and rutile polymorphs in a random assortment of TiO$_2$ nanoparticles and explore the potential for using this to study polymorph-specific bandgaps, composition, and crystal structure / orientation. We conclude with a discussion of how to interpret these results and emphasize that this workflow is capable of addressing an enormous variety of nanoscale materials challenges in a broad range of systems.

\section*{Methods}

\subsection*{Data acquisition}

A commercial powder of TiO$_2$ (Degussa P25) was dispersed in isopropanol, ultrasonicated, and dropped onto a holey carbon grid.  This powder has a nominal mixture of approximately 67\% anatase and 33\% rutile nanoparticles and is commonly used due to its high photocatalytic activity.  Contamination effects were reduced by heating the grid under an infrared lamp for 30 minutes prior to insertion in the TEM.  A larger agglomerate of multiple titania nanoparticles was identified and scanned twice to acquire scanning diffraction (known as 4D-STEM) as well as low-loss EELS, core-loss EELS, and EDX.  The scanned region was constrained to be approximately the same in both scans.

The TEM used for these experiments is a double aberration-corrected Themis Z by Thermo Fischer.  The instrument was operated at 300 kV and aberrations in the condenser system lenses were corrected up to 5th order.  The TEM was first configured for nanobeam diffraction with a probe current of 10~pA and a semi-convergence angle of 0.46~mrad.  The agglomerate was scanned with this configuration using a custom-written software package designed to collect a diffraction pattern at each probe position on a OneView camera (Gatan Inc.).  The acquisition speed was approximately 300 samples per second and the scan grid was $110 \times 190$ pixels.  

Following scanning diffraction, the TEM was configured for monochromated STEM-EELS using a convergence angle of 21.4~mrad, a collection angle of 23~mrad, and an estimated probe current of 150~pA.  With this electron optical configuration and using an energy dispersion of 0.01 eV / channel, the monochromater is capable of achieving an energy resolution of 210~meV as measured by taking the full-width at half-maximum of a rapidly-acquired zero-loss peak (ZLP).  The probe was scanned at a rate of approximately 180 pixels per second with an EDX, low-loss EELS, and core-loss EELS spectrum simultaneously acquired at each probe position.  The magnetic prism was continuously modified with a sawtooth waveform in order to shift the position of both EELS spectra on the spectrometer charge-coupled display in order to perform gain averaging \cite{bosman_optimizing_2008}. Approximately the same region was scanned in both cases.

\subsection*{Data analysis}

The EELS data were first treated by removing spectral artefacts and aligning the spectra using the energy offset calculated from the low-loss EELS datacube using custom-written MATLAB code.  Noise homoscedasticity was enforced by subjecting the core-loss EELS data to the Generalized Anscombe Transformation \cite{makitalo_exact_2013,makitalo_optimal_2013} prior to normalization and data fusion.  Data fusion, datablock decomposition, and geometric unmixing was performed following the workflow outlined in Thersleff et al. \cite{thersleff_dissecting_2020}.  Additional steps needed for spatial registration are described in the results section.

\section*{Acknowledgements}

T.~T.~acknowledges funding from the Swedish Research Council (project nr. 2016-05113).  C.-W.~T.~and T.~T.~acknowledge funding from the Swedish Strategic Research foundation (project nr. ITM17-0301).

\section*{Author contributions statement}

T.~T.~conceptualized the "hypermodal data fusion" workflow, designed the experiment, acquired the data, wrote the code, performed the data analysis, and authored the paper. C.~-W.~T assisted with the interpretation of the results and provided scientific feedback during manuscript writing.

\section*{Additional information}

\subsection*{Competing Interests}
The authors declare no competing interests.

\end{document}